\documentstyle[preprint,aps,epsfig]{revtex}
\begin{document}
\draft
%%%%%%%%%%%% Begin Cover Page %%%%%%%%%%%%%%%%%%%%%%%%%%%%%%%%%%%%%%%%%%
\preprint{\hfill ANL-HEP-PR-98-48}
\title{Scale Dependence of Squark and Gluino \\
Production Cross Sections}
\author{Edmond L. Berger$^a$, Michael Klasen$^a$, and Tim Tait$^{a,b}$}
\address{$^a$High Energy Physics Division,
             Argonne National Laboratory \\
             Argonne, Illinois 60439 \\
	$^b$Michigan State University, East Lansing, Michigan 48824}
\date{\today}
\maketitle

\begin{abstract} 
We investigate the choice of the renormalization and factorization scales 
for the production of squarks and gluinos at the Fermilab 
Tevatron collider as a function of the produced sparticle masses.  We 
determine a scale to be used in leading-order perturbative QCD calculations 
{\em optimized} such that the normalization of the next-to-leading order cross
section is reproduced.  Use of this optimal scale permits at least a partial 
implementation of next-to-leading order contributions in leading order Monte 
Carlo simulations.  We provide next-to-leading order predictions of the 
production cross sections for pairs of supersymmetric particles at the 
hadronic center-of-mass energy 2~TeV.
%The results are presented for three different choices of
%the scale in the next-to-leading order calculation which corresponds to
%the remaining theoretical error at this order of perturbative quantum
%chromodynamics. The corresponding cross section expectations can thus
%be chosen to be either optimistic, realistic, or pessimistic.
\end{abstract}
\vspace{0.2in}
%%%%%%%%%%%% End of Cover Page %%%%%%%%%%%%%%%%%%%%%%%%%%%%%%%%%%%%%%%%%

%%%%%%%%%%%%%% Begin Section I %%%%%%%%%%%%%%%%%%%%%%%%%%%%%%%%%%%%%%%%%
\section{Introduction}
\label{sec:1}
The possibility of supersymmetry (SUSY) at the electroweak scale and 
the ongoing search for the Standard Model (SM) Higgs boson constitute two 
major and related aspects of the motivation for the Tevatron upgrade currently 
under construction at Fermilab.  The increase in the center-of-mass energy to 
2~TeV and the luminosity to an expected 2 fb$^{-1}$, together with detector
improvements, should permit discovery or exclusion of supersymmetric partners 
of the standard model particles up to much higher masses than at present. A 
hadron collider like the Tevatron is particularly well suited for the 
production of strongly interacting sparticles, squarks and gluinos, among 
which the light stop eigenstate is expected to have the lowest mass value.

Experimental searches for supersymmetry rely heavily on Monte Carlo 
simulations of cross sections and event topologies. Two Monte Carlo generators 
in common use include SUSY processes; they are ISAJET \cite{Paige}
and SPYTHIA \cite{Sjostrand,Mrenna}.  Production cross sections may also 
be computed analytically from fixed-order quantum chromodynamics (QCD) 
perturbation theory.  Calculations that include contributions through 
next-to-leading order (NLO) in QCD have been performed for the production of 
squarks and gluinos \cite{Hopker}, stops \cite{Plehn}, and more recently of
sleptons \cite{Baer} and neutralinos \cite{Klasen}. The cross sections can
be calculated  as functions of the sparticle masses and thus do not depend
on a particular SUSY breaking mechanism.

Both the Monte Carlo approach and the fixed order approach have different
advantages and limitations.  Next-to-leading order perturbative calculations 
depend on very few parameters, e.g., the renormalization and factorization 
scales, and the dependence of the production cross sections on these 
parameters is reduced significantly in NLO with respect to leading order (LO). 
Therefore, the normalization of the cross section can be calculated quite 
reliably if one includes the NLO contributions.  On the other hand, the 
existing next-to-leading order calculations provide predictions only for 
fully inclusive quantities, e.g., a differential cross section for 
production of a squark or a gluino, after integration over all other particles 
and variables in the final state.  In addition, they do not include sparticle 
decays. This approach does not allow for event shape studies nor for 
experimental selections on missing energy or other variables associated with 
the produced sparticles or their decay products that are crucial if one wants 
to enhance the SUSY signal in the face of substantial backgrounds from 
Standard Model processes.

The natural strength of Monte Carlo simulations consists in the fact that
they generate event configurations that resemble those observed in experimental 
detectors.  Through their parton showers, these generators include, in the 
collinear approximation, contributions from all orders of perturbation theory.  
In addition, they incorporate phenomenological hadronization models, a 
simulation of particle decays, the possibility to implement experimental cuts, 
and event analysis tools.  However, the hard-scattering matrix elements in 
these generators are accurate only to leading order in QCD, and, owing to 
the rather complex nature of infra-red singularity cancellation in higher 
orders of perturbation theory, it remains a difficult challenge to incorporate 
the full structure of NLO contributions successfully in Monte Carlo simulations.
The limitation to leading-order hard-scattering matrix elements leads to 
large uncertainties in the normalization of the cross section.  The parton 
shower and hadronization models rely on tunable parameters, another source of 
uncertainties.

The aim of this paper is to improve the accuracy of the normalization of 
cross sections computed through Monte Carlo simulations.  We introduce some 
aspects of the reliability and normalization of next-to-leading order 
calculations while preserving the flexibility and event shape versatility 
of Monte Carlo simulations. This goal can be approached in at least two ways, 
with different secondary consequences.  The first method is simply to multiply  
the leading order cross section computed in the Monte Carlo simulation by an 
overall
\mbox{$K$-factor}.  The second, the method we adopt and investigate, is to 
select a renormalization and factorization (hard) scale in the Monte Carlo 
LO calculation such that, with this choice of hard scale, the normalization 
of the Monte Carlo LO calculation agrees with that of the NLO 
perturbative calculation.  The answer one obtains in both approaches will in 
general depend on which partonic subprocess one is considering and on the 
kinematics.   However, only a change in the hard scale will affect both the 
hard matrix element {\em and} the initial-state and final-state parton shower 
radiation.   A rescaling of the cross section by an overall $K$-factor will 
have no bearing on the parton shower radiation.  A reduction in the hard scale 
leads generally to less evolution and less QCD radiation, and vice-versa, in 
the initial- and final-state showering.  A change of the hard scale will be 
reflected in the normalization of the cross section as well as in the event 
shape.  We suggest that the approach we adopt is a  
%One can therefore hope that this avoids double counting of the first real
%emission and leads to a 
more consistent combination of next-to-leading order and parton shower effects. 

The outline of this paper is as follows: In Section 2, we briefly
review the calculation of next-to-leading order hadronic cross sections for
squark and for gluino production. In Section 3, we discuss the hard scale
dependence at leading order and at next-to-leading order.  Our main results
are presented in Section 4. There we show the dependence of the optimal
scale choices on the produced sparticle masses. A summary
is given in Section 5.

%%%%%%%%%%%%%% End of Section I %%%%%%%%%%%%%%%%%%%%%%%%%%%%%%%%%%%%%%%%%

%%%%%%%%%%%%%% Begin Section II %%%%%%%%%%%%%%%%%%%%%%%%%%%%%%%%%%%%%%%%
\section{Squark and Gluino Production at Next-to-Leading Order}
\label{sec:2}

We consider the pair production of strongly interacting
supersymmetric particles in proton-antiproton scattering. The total
cross section can be calculated as a function of the hadronic
center-of-mass energy $\sqrt{s_H}$ and the produced sparticle mass
$\tilde{m}$ through the factorization theorem
\begin{equation}
 \sigma^{\rm total}_{p\bar{p}} (s_H,\tilde{m}^2) = \hspace*{-5mm}
 \sum_{a,b=q,\bar{q},g}
 \int\mbox{d}x_a\mbox{d}x_b f^a_p (x_a,M^2) f^b_{\bar{p}} (x_b,M^2)
 \hat{\sigma}_{ab}(x_ax_bs_H,\tilde{m}^2;M^2).
\end{equation}
We set $\sqrt{s_H} = 2$ TeV for Tevatron Run II conditions.

The parton densities $f^{a,b}_{p,\bar{p}}$ depend on the
longitudinal light-cone momentum fractions $x_{a,b}$ of the quarks and gluons
in the proton and antiproton, respectively, as well as on the
factorization scale $M$. We treat the gluon and the five light quark
flavors as massless and use the CTEQ4M parametrization \cite{CTEQ4}
throughout this paper. For the top quark, we assume a mass of 175~GeV
and neglect its contribution to the parton densities.

The partonic cross section $\hat{\sigma}$ depends on the partonic
center-of-mass energy $x_ax_bs_H$, on the produced sparticle mass
$\tilde{m}$, and on the factorization scale $M$. It can be calculated
in fixed order perturbative QCD by an expansion in the strong coupling 
strength 
\begin{equation}
 \hat{\sigma}_{ab}(x_ax_bs_H,\tilde{m}^2;M^2) = \alpha_s^2(\mu^2)
 \hat{\sigma}^{\rm tree}_{ab} + \alpha_s^3(\mu^2)
 \hat\sigma^{\rm loop}_{ab} (\mu^2)
 +~\alpha_s^3(\mu^2) \hat\sigma^{\rm real}_{ab}(M^2) + 
 {\cal O} (\alpha_s^4) .
\end{equation}
The renormalization scale is denoted $\mu$. At leading order, the only
dependence on $\mu$ arises in the strong coupling strength $\alpha_s$
since only tree-level $2\rightarrow 2$ processes are taken into account.
The partonic subprocesses that contribute at this stage are:
\begin{eqnarray}
 q + \bar{q}, g + g & \longrightarrow & \tilde{q}   + \bar{\tilde{q}} ,  \\
 q + q              & \longrightarrow & \tilde{q}   +      \tilde{q} ,    \\
 q + \bar{q}, g + g & \longrightarrow & \tilde{g}   +      \tilde{g} ,   \\
 q + g              & \longrightarrow & \tilde{q}   +      \tilde{g} ,   \\
 q + \bar{q}, g + g & \longrightarrow & \tilde{t}_1 + \bar{\tilde{t}}_1 .
\end{eqnarray}

At next-to-leading order, both virtual loop diagrams and real  
emission diagrams contribute.  The real emission contributions are integrated 
over the soft and collinear regions of the additional third parton. Through 
the renormalization procedure, logarithmic terms arising in the virtual 
diagrams partly cancel the leading order renormalization scale dependence.
The collinear singularities in the real emission diagrams
are absorbed into the parton densities, introducing logarithmic dependence on 
the factorization scale.  Therefore, we also expect a reduced
factorization scale dependence.  A detailed presentation of the
next-to-leading order calculation can be found in Refs.~\cite{Hopker,Plehn}.

In this paper, all sparticle masses are set to
\begin{equation}
 \tilde{m} = [m_{\tilde{q}};m_{\tilde{g}};(m_{\tilde{q}}
               +m_{\tilde{g}})/2] 
           = 250 ~{\rm GeV},
\end{equation}
if not stated otherwise,
with the exception of the light stop mass $m_{\tilde{t}_1}=153$~GeV.
The strong Yukawa coupling between the top quark, its supersymmetric
partner, and the Higgs field leads to large mixing of the left- and
right-handed stops and to a $\tilde{t}_{1}(\tilde{t}_{2})$ mass that is 
expected to lie
below (above) the other squark masses. Therefore, we treat the light
stop $\tilde{t}_1$ here separately from the other squarks and ignore
the heavy stop $\tilde{t}_2$. The particular mass value of
$m_{\tilde{t}_1}=153$~GeV
results from a minimal supergravity-inspired solution of the
renormalization group equations with input values $m_0 = m_{1/2}
= 100$~GeV, $A_{0} = 300$~GeV, $\tan\beta = 1.75$ and $\mu > 0$. The
mixing angle
is found to be $\sin(2\tilde{\theta}) = - 0.99$. This is the only place
where an assumption of a specific SUSY breaking mechanism is made. The
squark and gluino mass values are chosen in such a way that they
are above experimental exclusion limits but within reach of Run II at
the Tevatron.

\begin{figure}
 \begin{center}
  {\unitlength1cm
  \begin{picture}(12,17)
   \epsfig{file=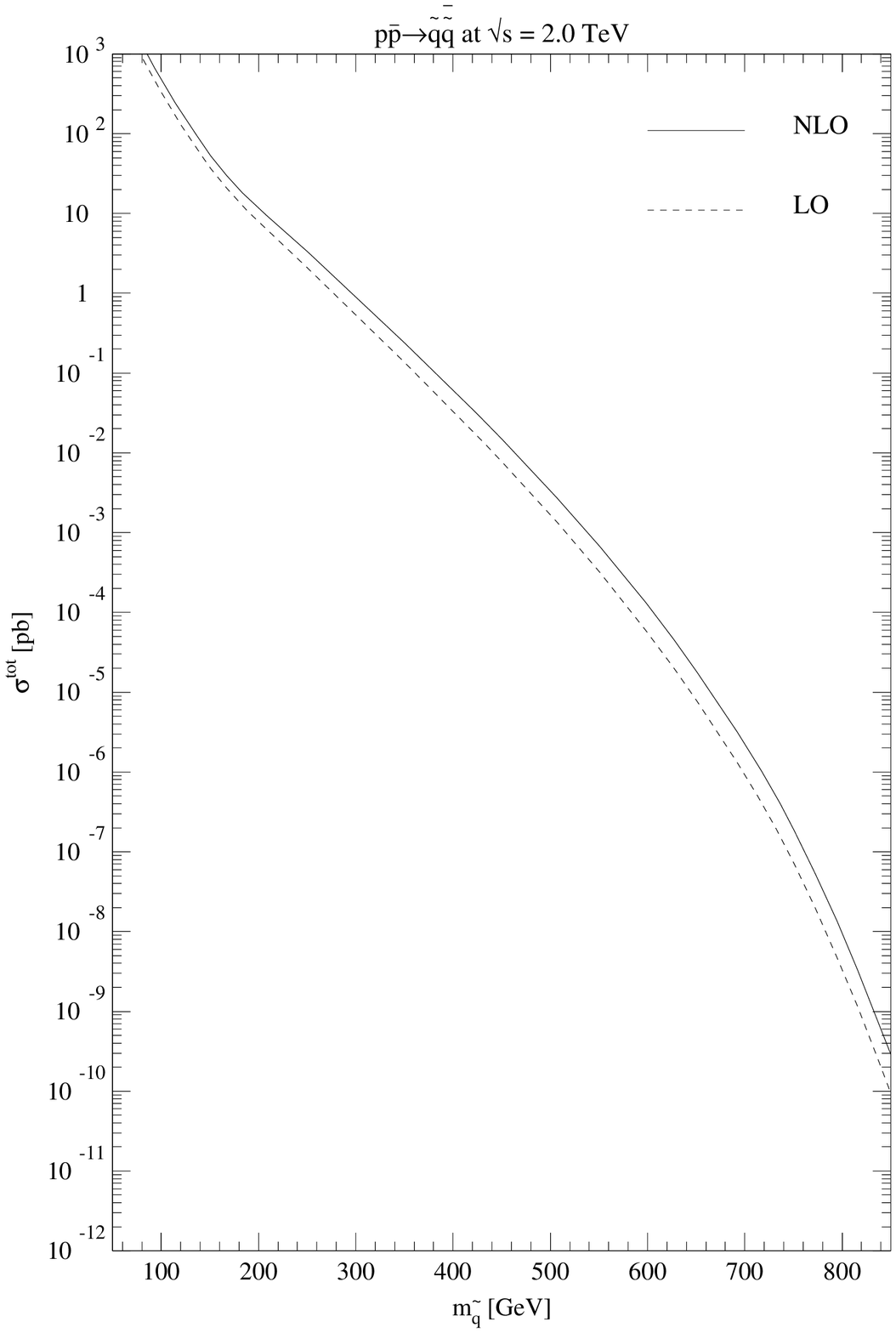,bbllx=60pt,bblly=95pt,bburx=485pt,bbury=730pt,%
           height=17cm,clip=}
   \put(-7.4,13.6){\vector(0,1){0.2}         }
   \put(-7.6,14.1){+ 59 \% }
   \put(-7.4,13.6){\vector(1,0){0.2}         }
   \put(-7.0,13.6){+ 19 GeV}
  \end{picture}}
 \end{center}
\caption{Total cross section for squark-antisquark production at Run II
of the Tevatron as a function of squark mass. The next-to-leading order
cross section (full curve) exceeds the leading order cross section
(dashed curve) by 59~\%. Alternatively, for any measured cross
section the corresponding squark mass is shifted by 19~GeV.
The gluino mass has been fixed at $m_{\tilde{g}} = 250$ GeV.}
\label{fig1a}
\end{figure}

\begin{figure}
 \begin{center}
  {\unitlength1cm
  \begin{picture}(12,17)
   \epsfig{file=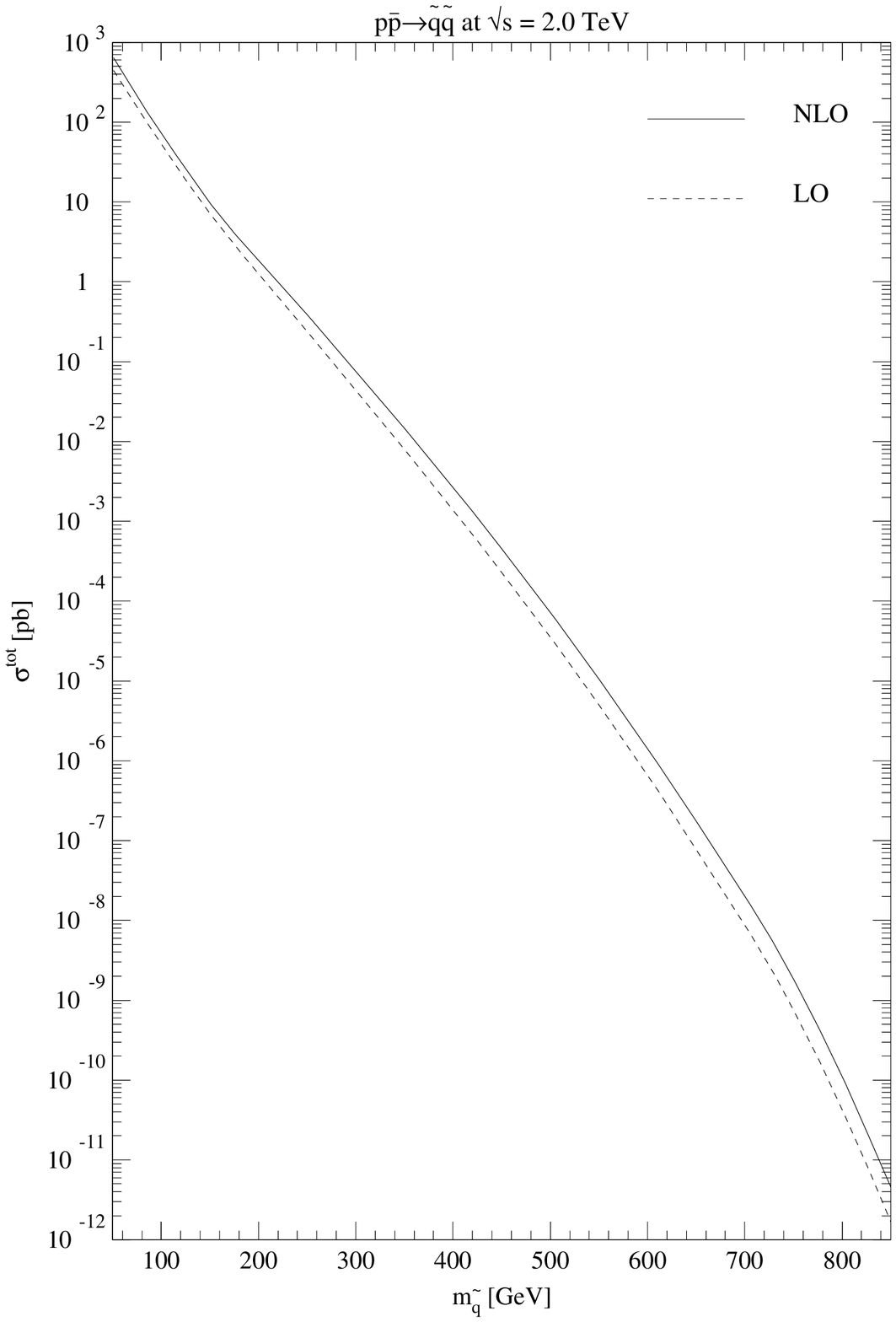,bbllx=60pt,bblly=95pt,bburx=485pt,bbury=730pt,%
           height=17cm,clip=}
   \put(-7.4,12.6){\vector(0,1){0.2}         }
   \put(-7.6,13.1){+ 64 \% }
   \put(-7.4,12.6){\vector(1,0){0.2}         }
   \put(-7.0,12.6){+ 17 GeV}
  \end{picture}}
 \end{center}
\caption{Total cross section for squark pair production at Run II
of the Tevatron as a function of squark mass. The next-to-leading order
cross section (full curve) exceeds the leading order cross section
(dashed curve) by 64~\%. Alternatively, for any measured cross
section the corresponding squark mass is shifted by 17~GeV.
The gluino mass has been fixed at $m_{\tilde{g}} = 250$ GeV.}
\label{fig1b}
\end{figure}

\begin{figure}
 \begin{center}
  {\unitlength1cm
  \begin{picture}(12,17)
   \epsfig{file=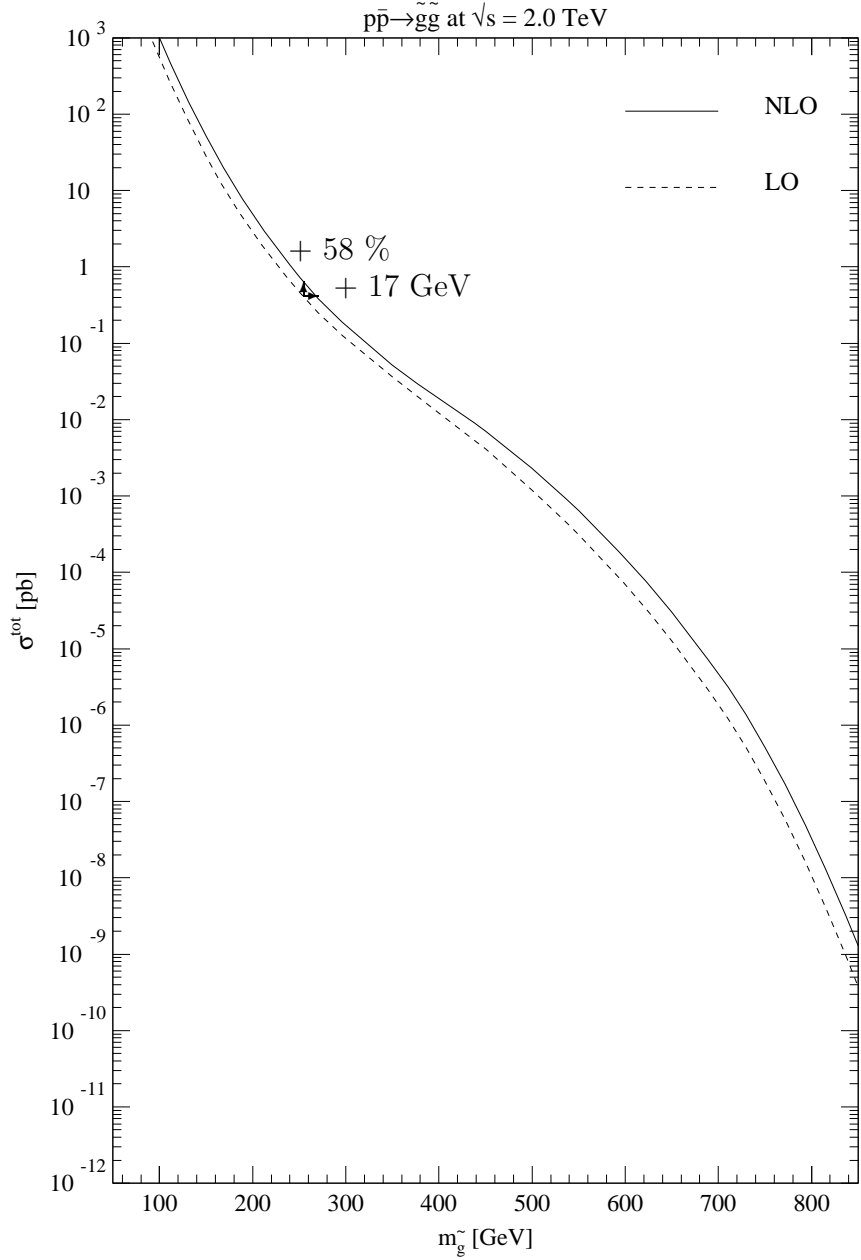,bbllx=60pt,bblly=95pt,bburx=485pt,bbury=730pt,%
           height=17cm,clip=}
   \put(-7.4,12.95){\vector(0,1){0.2}         }
   \put(-7.6,13.45){+ 58 \% }
   \put(-7.4,12.95){\vector(1,0){0.2}         }
   \put(-7.0,12.95){+ 17 GeV}
  \end{picture}}
 \end{center}
\caption{Total cross section for gluino pair production at Run II
of the Tevatron as a function of squark mass. The next-to-leading order
cross section (full curve) exceeds the leading order cross section
(dashed curve) by 58~\%. Alternatively, for any measured cross
section the corresponding gluino mass is shifted by 17~GeV.
The squark mass has been fixed at $m_{\tilde{q}} = 250$ GeV.}
\label{fig1c}
\end{figure}

\begin{figure}
 \begin{center}
  {\unitlength1cm
  \begin{picture}(12,17)
   \epsfig{file=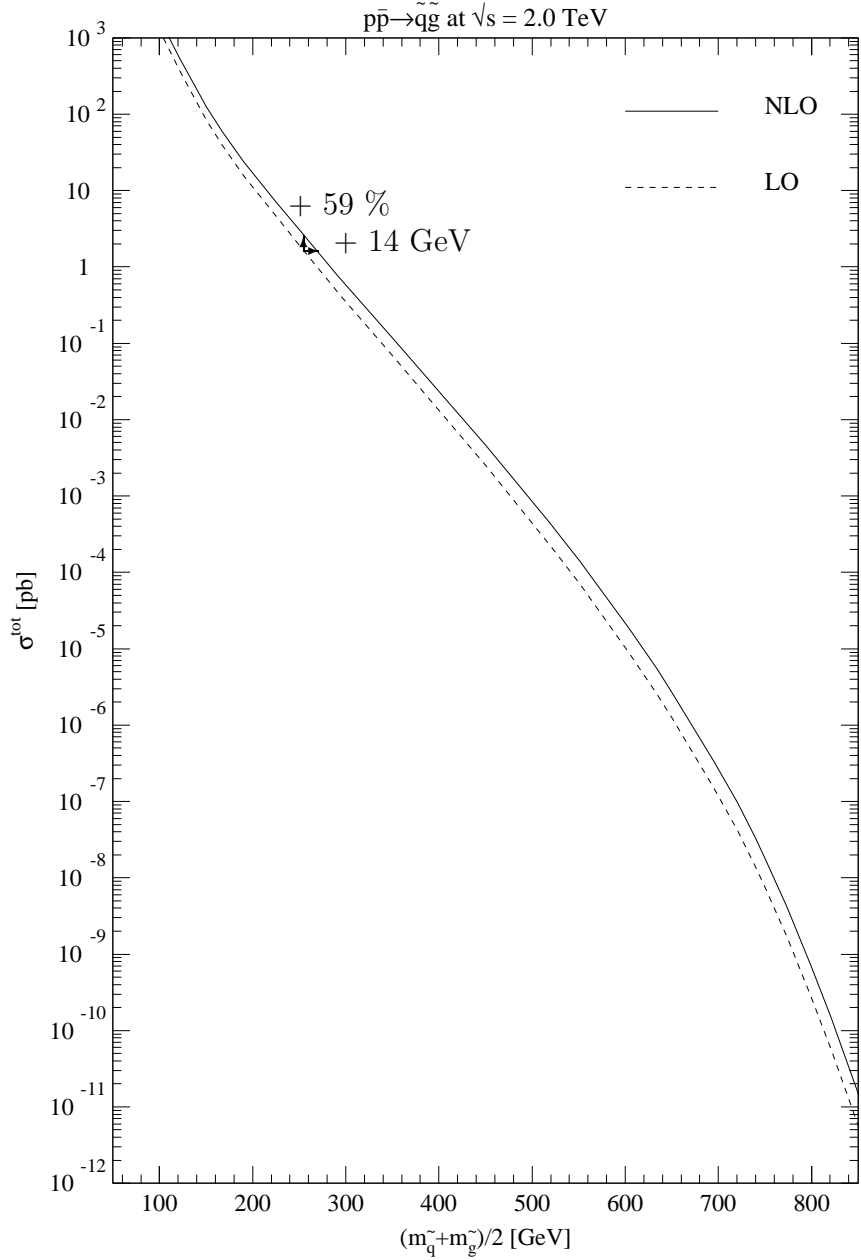,bbllx=60pt,bblly=95pt,bburx=485pt,bbury=730pt,%
           height=17cm,clip=}
   \put(-7.4,13.55){\vector(0,1){0.2}         }
   \put(-7.6,14.05){+ 59 \% }
   \put(-7.4,13.55){\vector(1,0){0.2}         }
   \put(-7.0,13.55){+ 14 GeV}
  \end{picture}}
 \end{center}
\caption{Total cross section for associated production of a squark and
gluino at Run II of the Tevatron as a function of the average squark and
gluino mass. The next-to-leading order cross section (full curve) exceeds
the leading order cross section (dashed curve) by 59~\%. Alternatively,
for any measured cross section the corresponding average squark and gluino
mass is shifted by 14~GeV. For this plot, we have set $m_{\tilde{q}} =
m_{\tilde{g}}$.}
\label{fig1d}
\end{figure}

\begin{figure}
 \begin{center}
  {\unitlength1cm
  \begin{picture}(12,17)
   \epsfig{file=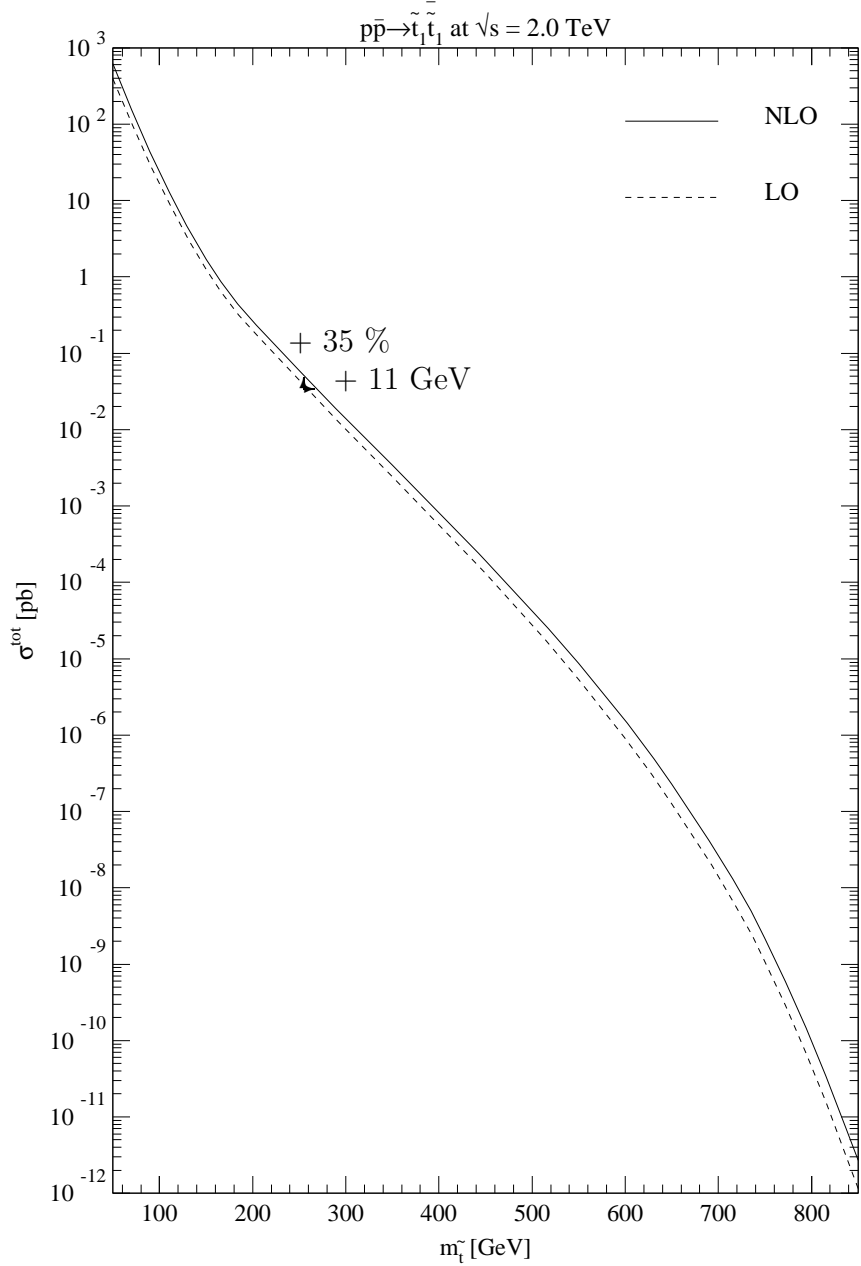,bbllx=60pt,bblly=95pt,bburx=485pt,bbury=730pt,%
           height=17cm,clip=}
   \put(-7.4,11.85){\vector(0,1){0.15}         }
   \put(-7.6,12.35){+ 35 \% }
   \put(-7.4,11.85){\vector(1,0){0.15}         }
   \put(-7.0,11.85){+ 11 GeV}
  \end{picture}}
 \end{center}
\caption{Total cross section for stop-antistop production at Run II
of the Tevatron as a function of the stop mass $m_{\tilde{t}_1}$.
The next-to-leading order
cross section (full curve) exceeds the leading order cross section
(dashed curve) by 35~\%. Alternatively, for any measured cross
section the corresponding stop mass is shifted by 11~GeV.
The light squark mass, the gluino mass, and the mixing parameter are
set to $m_{\tilde{q}}= 256$ GeV, $m_{\tilde{g}} = 284$ GeV, and
$\sin(2\tilde{\theta}) = - 0.99$.}
\label{fig1e}
\end{figure}

In Figs.~\ref{fig1a}-\ref{fig1e} we present the next-to-leading order
values for the  production cross sections for all strong SUSY channels
as a function of sparticle mass at 2 TeV. Also provided are the leading
order values. 
In obtaining the numerical results presented in this paper, we use
a modified version of the PROSPINO code \cite{Prospino}, with the FORTRAN
altered to run on the VAX/VMS platform.  We have verified that our results
agree with those in Refs. \cite{Hopker} and \cite{Plehn} when we select the
same center-of-mass energy and other parameters.
For our results, we set the common hard scale $\mu = M$ equal to 
$\mu = {\tilde{m}}$ in both the leading and the next-to-leading order 
calculations.  
For our next-to-leading order results, we use the hard matrix elements 
through next-to-leading order, parton densities evolved with next-to-leading 
order Altarelli-Parisi kernels, and the two-loop expression for the strong 
coupling strength $\alpha_s(\mu)$.  For the leading order results, the only 
change we make is 
to use the leading order hard matrix element.  In this way, we are able to 
identify how much of the increase in cross section at next-to-leading order 
is due to the hard scattering matrix element.  For the leading order results,
one could also employ leading order definitions of all three components.  
It could be argued that our procedure might not be consistent philosophically. 
However, the difference is of next-to-leading order so that both 
determinations of the leading order cross section are equally consistent up to 
terms of ${\cal O} (\alpha_s^3)$.  Of more importance to us is to study the 
effect of the next-to-leading order matrix element and not the well-known 
universal effects of the higher order terms in the parton densities or the 
coupling constant. In addition, the extraction of leading order parton 
densities is more uncertain, and some packages of parton densities (e.g., 
MRS \cite{MRS}) do not include a leading order option.

In the case of SUSY particle production, the next-to-leading order 
contributions are known to increase the production cross sections 
by 50~\% and more.  For the various channels we consider, the increase can 
be seen explicitly in Figs.~\ref{fig1a}-\ref{fig1e}. 
For example, in Fig.~\ref{fig1a} we plot the total cross section for
squark-antisquark production as a function of the squark mass. The 
next-to-leading order cross section (full curve) lies above the leading 
order cross section (dashed curve) by 59~\%.  This increase translates into a 
shift in the lower limit of the produced squark mass of 19~GeV.
The cross sections for gluino pair production (Fig.~\ref{fig1c}) and the
associated production of squarks and gluinos of equal mass (Fig.~\ref{fig1d})
are of similar magnitude, whereas the squark pair production
(Fig.~\ref{fig1b}) and stop-antistop production (Fig.~\ref{fig1e}) cross
sections are smaller by about an order of magnitude \cite{Hopker,Plehn}.

%%%%%%%%%%%%%% End of Section II %%%%%%%%%%%%%%%%%%%%%%%%%%%%%%%%%%%%%%%%%

%%%%%%%%%%%%%% Begin Section III %%%%%%%%%%%%%%%%%%%%%%%%%%%%%%%%%%%%%%%%
\section{Scale Dependence at Leading Order and at Next-to-leading Order}
\label{sec:3}

In this section, we study the dependence of the leading and next-to-leading
order cross sections for squark and gluino production on the renormalization
and factorization scales. The cross section is expected to depend principally 
on the produced sparticle mass $\tilde{m}$. Therefore, we set the
renormalization and factorization scales equal, $\mu = M \equiv Q$, and
calculate our results as a function of $Q/{\tilde{m}}$.

\begin{figure}
 \begin{center}
  {\unitlength1cm
  \begin{picture}(12,17)
   \epsfig{file=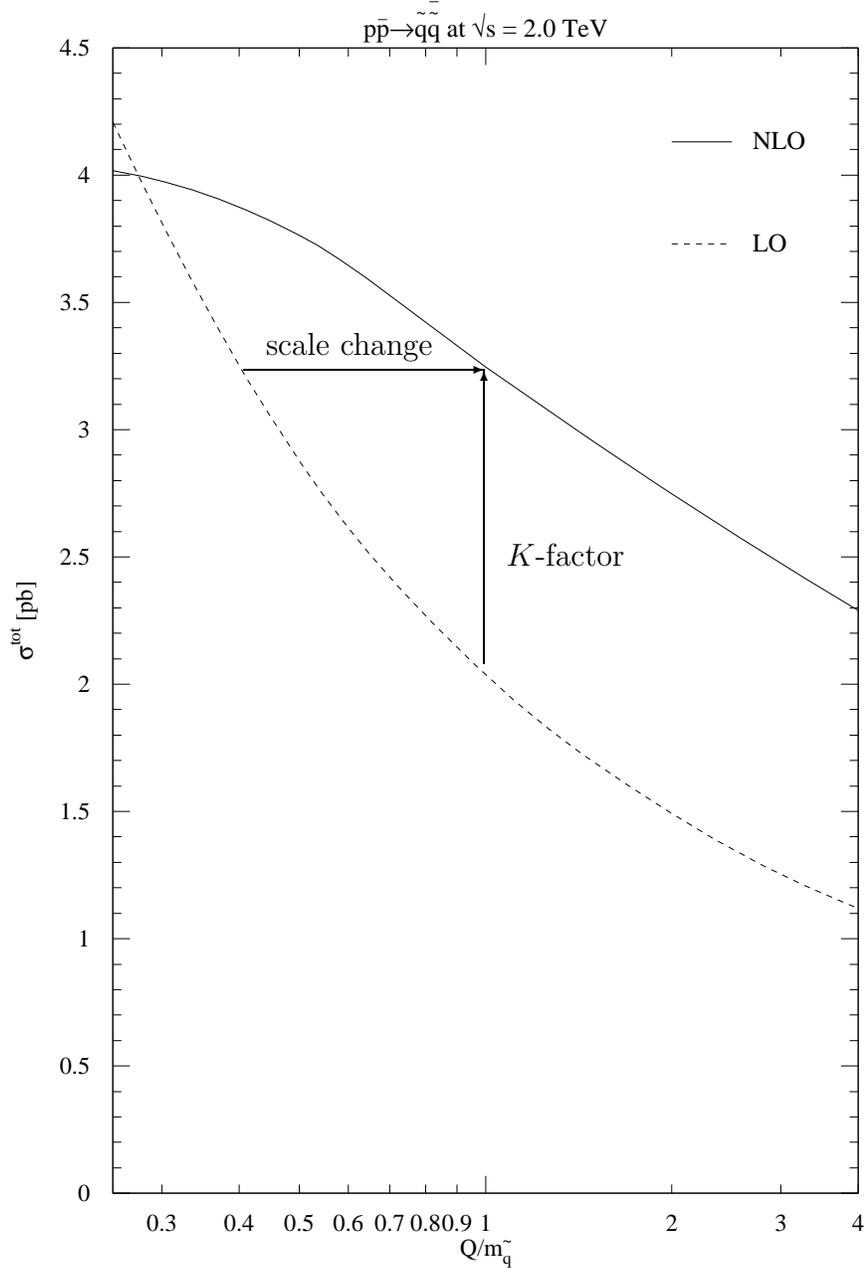,bbllx=60pt,bblly=95pt,bburx=485pt,bbury=730pt,%
           height=17cm,clip=}
   \put(-5  ,8.2){\vector(0,1){3.9}}
   \put(-4.7,9.5){$K$-factor}
   \put(-8.2,12.1){\vector(1,0){3.2}}
   \put(-7.9,12.3){scale change}
  \end{picture}}
 \end{center}
\caption{Total cross section for squark-antisquark production at Run II
of the Tevatron as a function of the common renormalization and
factorization scale $Q$. The squark mass has been fixed at $m_{\tilde{q}}=
250$~GeV, and $m_{\tilde{g}}= 250$ GeV. For
$Q=m_{\tilde{q}}$, the next-to-leading order cross section
(full curve) can be obtained from the leading order cross section
(dashed curve) if the latter is multiplied by a $K$-factor of 1.59.
The same result is obtained if a hard scale of $Q = 0.41~m_{\tilde{q}}$ 
is used in the calculation of the leading order cross section.}
\label{fig2}
\end{figure}

Figure \ref{fig2} presents the total cross section for squark-antisquark
production as a function of the common renormalization and
factorization scale $Q$. The squark mass has been fixed at $m_{\tilde{q}}=
250$~GeV. It is common to estimate the theoretical uncertainty by examining 
the variation of the calculated cross section over an interval of 
$Q/m_{\tilde{q}} = [0.5;2.0]$.  The strong scale dependence of 63~\% for the 
leading order cross section (dashed curve) is reduced to 31~\% (i.e., 
$\pm 15.5$~\% about a central value) for the 
next-to-leading order cross section (full curve), smaller but still 
considerable.  For $Q=m_{\tilde{q}}$, the next-to-leading order cross section 
can be obtained from the leading order cross section if the latter is 
multiplied by a $K$-factor of 1.59.  The same increase is obtained if a scale 
$Q = 0.41~m_{\tilde{q}}$ is used in the calculation of the leading order 
cross section.

\begin{figure}
 \begin{center}
  {\unitlength1cm
  \begin{picture}(12,17)
   \epsfig{file=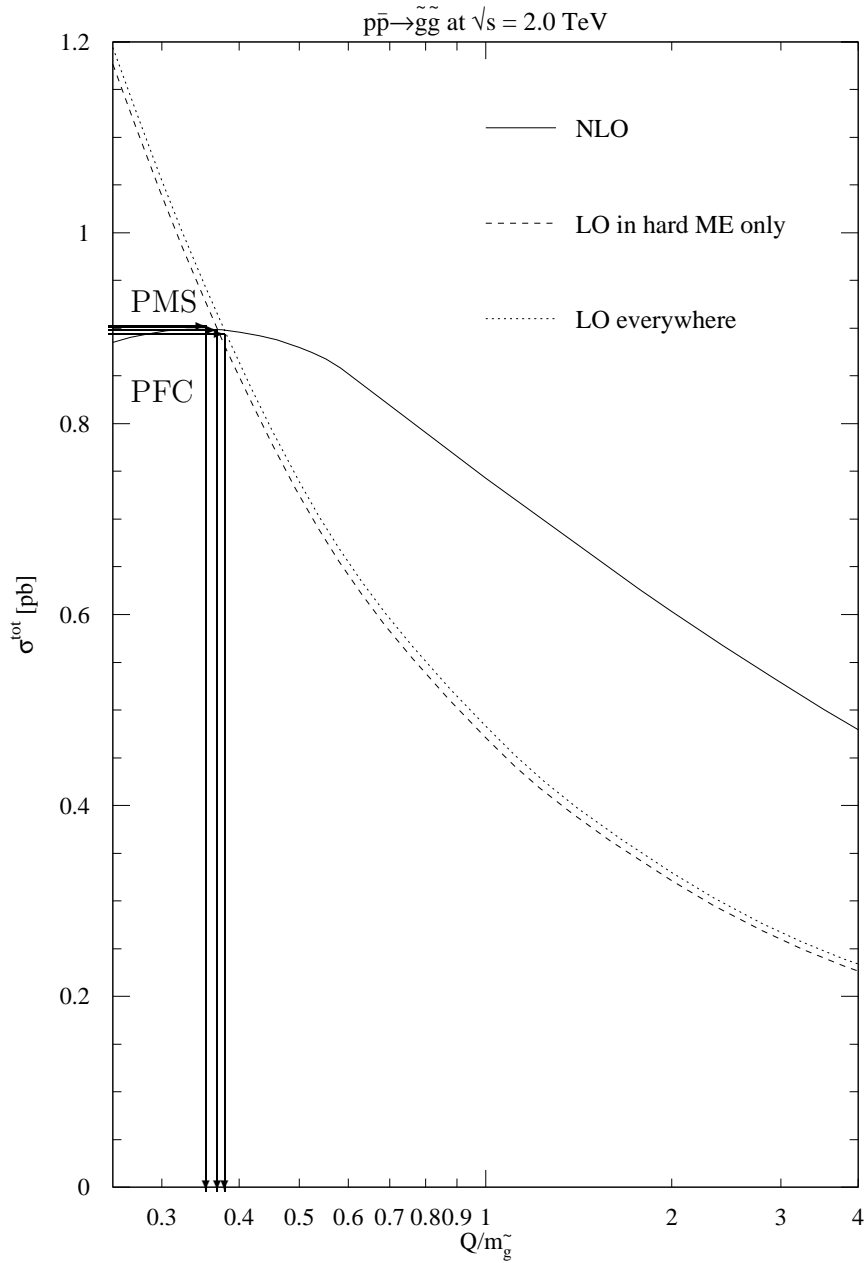,bbllx=60pt,bblly=95pt,bburx=485pt,bbury=730pt,%
           height=17cm,clip=}
   \put(-10  ,12.6 ){\vector(1, 0){ 1.3}}
   \put(-8.7 ,12.6 ){\vector(0,-1){11.5}}
   \put(-10  ,12.55){\vector(1, 0){ 1.45}}
   \put(-8.55,12.55){\vector(0,-1){11.45}}
   \put(-10  ,12.5 ){\vector(1, 0){ 1.55}}
   \put(-8.45,12.5 ){\vector(0,-1){11.4 }}
   \put(-9.7 ,12.8 ){PMS}
   \put(-9.7 ,11.6 ){PFC}
  \end{picture}}
 \end{center}
\caption{Total cross section for gluino pair production at Run II
of the Tevatron as a function of the common renormalization and
factorization scale $Q$. The gluino mass has been fixed at $m_{\tilde{g}}=
250$~GeV, and $m_{\tilde{q}}= 250$ GeV.
The scales preferred by the Principle of Minimal Sensitivity (PMS,
$Q = 0.35~m_{\tilde{g}}$) and the Principle of Fastest Convergence (PFC,
$Q = 0.36~m_{\tilde{g}}$) are close to each other if the use of leading 
order expressions is restricted to the hard matrix element. If the parton 
densities and $\alpha_s$ are calculated in leading order as well, the PFC 
scale shifts to $Q = 0.37~m_{\tilde{g}}$.}
\label{fig3}
\end{figure}

\begin{figure}
 \begin{center}
  {\unitlength1cm
  \begin{picture}(12,17)
   \epsfig{file=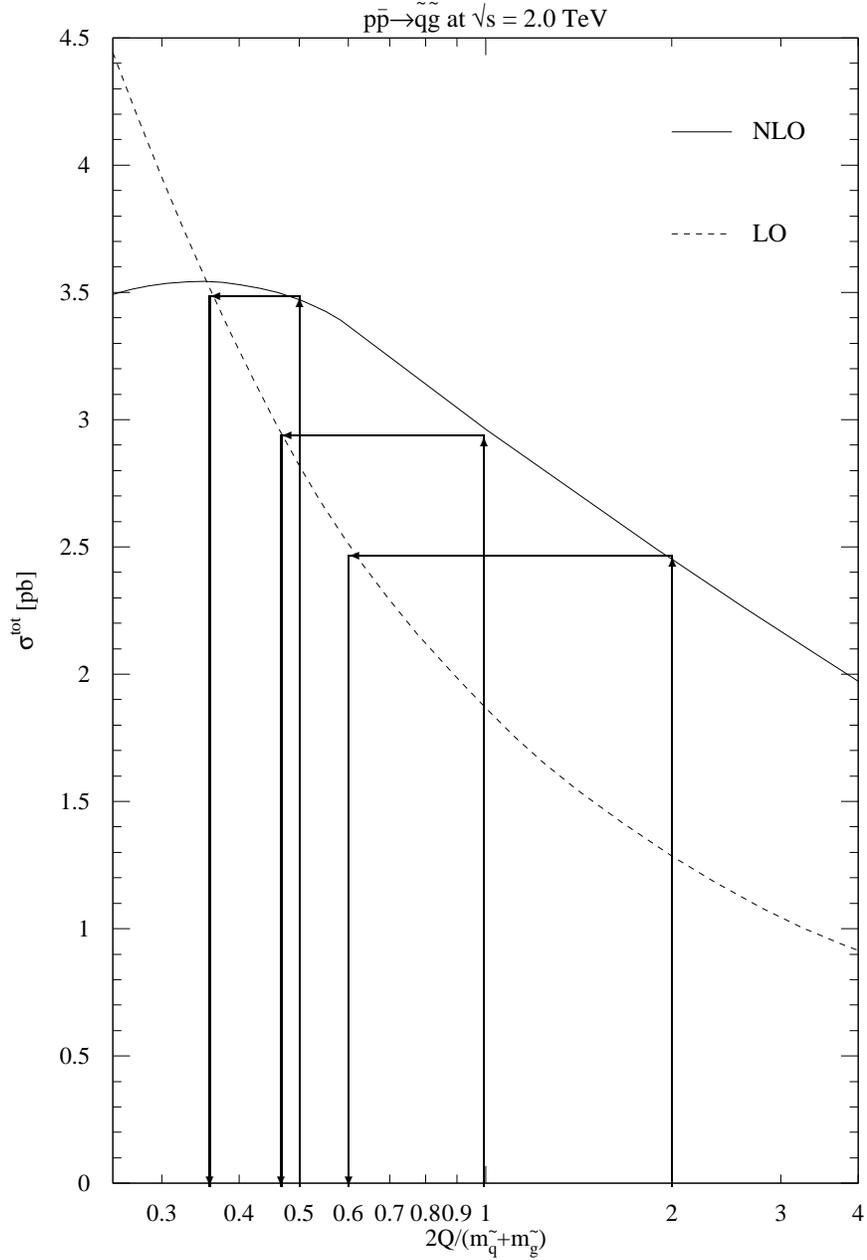,bbllx=60pt,bblly=95pt,bburx=485pt,bbury=730pt,%
           height=17cm,clip=}
   \put(-7.45, 1.1){\vector( 0, 1){11.85}}
   \put(-7.45,12.95){\vector(-1, 0){ 1.2}}
   \put(-8.65,12.95){\vector( 0,-1){11.85}}
   \put(-5. , 1.1){\vector( 0, 1){10  }}
   \put(-5. ,11.1){\vector(-1, 0){ 2.7}}
   \put(-7.7,11.1){\vector( 0,-1){10  }}
   \put(-2.5, 1.1){\vector( 0, 1){ 8.4}}
   \put(-2.5, 9.5){\vector(-1, 0){ 4.3}}
   \put(-6.8, 9.5){\vector( 0,-1){ 8.4}}
  \end{picture}}
 \end{center}
\caption{Total cross section for associated production of a squark and gluino 
at Run II of the Tevatron as a function of the common renormalization and
factorization scale $Q$. The squark and gluino masses have been fixed at
$m_{\tilde{q}} = m_{\tilde{g}} = 250$~GeV. The three different scale choices of
$2Q_{\rm NLO}/(m_{\tilde{q}}+m_{\tilde{g}}) = [0.5;1.0;2.0]$ define the 
theoretical uncertainty at next-to-leading order. The corresponding cross 
sections are reproduced in leading order by 
$2Q_{\rm LO}/(m_{\tilde{q}}+m_{\tilde{g}}) = [0.35;0.475;0.6]$.}
\label{fig4}
\end{figure}

In Figs.~\ref{fig3} and \ref{fig4}, we show the hard scale dependence of the 
cross sections for gluino pair production and associated production of a 
squark and gluino.  In these plots, as in those for squark-squark and 
stop-antistop production, a scale variation of about $\pm 18$~\% about a 
central value is observed in the values of the next-to-leading order cross 
sections, similar to that in Fig.~\ref{fig2}.

In the literature, two scale choices are sometimes preferred.  We advocate 
neither of these, but we mention them to distinguish them from the 
phenomenologically-based range of scales that we suggest in Sec.~IV be used 
in leading order calculations.  The Principle of Minimal Sensitivity (PMS) 
scale is specified by \cite{Stevenson}
\begin{equation}
 \frac{{\rm d}\sigma_{\rm NLO}}{{\rm d}Q} (Q_{\rm PMS}) = 0.
\end{equation}
It has the advantage that in the neighborhood of $Q_{\rm PMS}$, a
variation in the scale results in no change of the next-to-leading 
order (NLO) cross
section. The Principle of Fastest Convergence (PFC) scale is defined by
\cite{Grunberg}
\begin{equation}
 \sigma_{\rm NLO} (Q_{\rm PFC}) = \sigma_{\rm LO} (Q_{\rm PFC}).
\end{equation}
Its virtue is that the NLO correction is absent, and the
perturbative series seems to converge rapidly.  Although
next-to-next-to-leading (NNLO) order terms might well spoil both, 
the prescriptions provide some guidance until these NNLO
corrections are calculated.

The Principle of Minimal Sensitivity and the Principle of Fastest
Convergence have been shown to be consistent with each other up
to NNLO terms \cite{Kubo}. However, the Principle of Fastest
Convergence answer depends on whether the leading order (LO) cross section 
is defined with LO expressions throughout or in the hard matrix
element only. Due to the larger value of $\alpha_s$ in the one-loop
approximation, the first approach leads to a larger LO cross
section. This alternative usually moves the PFC scale away from 
the PMS scale.
 
In Fig.~\ref{fig3} we present the total cross section for gluino pair
production as a function of the common renormalization and
factorization scale $Q$. The gluino mass has been fixed at $m_{\tilde{g}}=
250$~GeV. The scales preferred by the Principle of Minimal Sensitivity (PMS),
$Q = 0.35~m_{\tilde{g}}$, and the Principle of Fastest Convergence (PFC),
$Q = 0.36~m_{\tilde{g}}$, are close to each other if the leading order choice 
is restricted to the hard matrix element only. If the parton densities and
$\alpha_s$ are calculated in leading order as well, the PFC scale shifts to
$Q = 0.37~m_{\tilde{g}}$.

%%%%%%%%%%%%%% End of Section III %%%%%%%%%%%%%%%%%%%%%%%%%%%%%%%%%%%%%%%%%

%%%%%%%%%%%%%% Begin Section IV %%%%%%%%%%%%%%%%%%%%%%%%%%%%%%%%%%%%%%%%
\section{Optimal Scale Choices}
\label{sec:4}

The aim of this section is to determine optimal scale choices to be 
used in leading order calculations of squark and gluino production at 
the Tevatron.  If implemented in leading order Monte Carlo generators 
such as ISAJET \cite{Paige} or SPYTHIA \cite{Sjostrand,Mrenna}, these optimal 
scales will reproduce the size of the NLO cross section:
\begin{equation}
 \sigma_{\rm  LO} (Q_{\rm  LO}/\tilde{m}) =
 \sigma_{\rm NLO} (Q_{\rm NLO}/\tilde{m}).
\end{equation}

To illustrate our definition of optimal scale choices, we show in
Fig.~\ref{fig4} the total cross section for associated squark and gluino
production as a function of the common renormalization and
factorization scale $Q$. The squark and gluino masses are fixed at
$m_{\tilde{q}} = m_{\tilde{g}} = 250$~GeV.  The NLO cross section depends 
on the scale choice.  Correspondingly, there is not a unique choice of 
scale in LO that will reproduce the NLO value of the cross section but, 
rather, a range of values that will reproduce the band of NLO values.  
As is done commonly, we define the theoretical uncertainty 
in the NLO value by the spread associated with three different scale choices
$2Q_{\rm NLO}/(m_{\tilde{q}}+m_{\tilde{g}}) = [0.5;1.0;2.0]$.
The corresponding cross sections are reproduced in leading order
by the somewhat narrower band of scales, 
$2Q_{\rm LO}/(m_{\tilde{q}}+m_{\tilde{g}}) = [0.35;0.475;0.6]$.

\begin{figure}
 \begin{center}
  {\unitlength1cm
  \begin{picture}(12,17)
   \epsfig{file=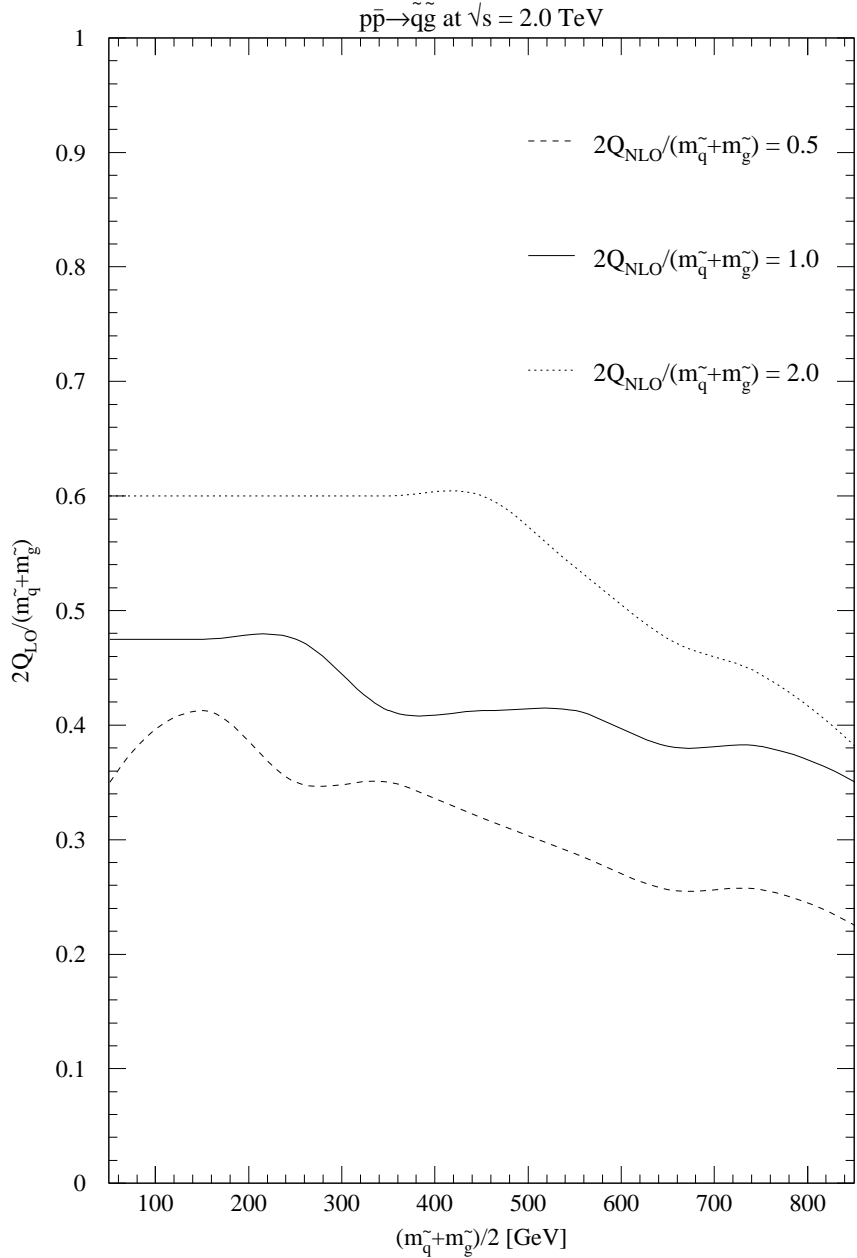,bbllx=60pt,bblly=95pt,bburx=485pt,bbury=730pt,%
           height=17cm,clip=}
  \end{picture}}
 \end{center}
\caption{Optimal scale choices for associated production of a squark and 
gluino at
Run II of the Tevatron as a function of the average squark and gluino mass
$(m_{\tilde{q}}+m_{\tilde{g}})/2$. For the central (full) curve, the NLO scale
has been chosen equal to the average squark and gluino mass. The dashed and
dotted curves correspond to the remaining uncertainty at NLO from a variation 
of $2Q_{\rm NLO}/(m_{\tilde{q}}+m_{\tilde{q}}) = [0.5;2.0]$.  For this plot,
we have set $m_{\tilde{q}}=m_{\tilde{g}}$.}
\label{fig5}
\end{figure}

The cross sections for the hadroproduction
of squarks and gluinos depend strongly on the produced sparticle mass. We
expect some sparticle mass dependence also in the optimal scale choices that
reproduce the size of the next-to-leading order cross sections at
leading order. This dependence is demonstrated in Fig.~\ref{fig5}, where we 
plot the optimal scale choices for associated squark and gluino 
production as a function of the average squark and gluino mass
$(m_{\tilde{q}}+m_{\tilde{g}})/2$. For the central (full) curve, the NLO scale
is chosen equal to the average squark and gluino mass. The dashed and
dotted curves define the remaining uncertainty at NLO from a variation of
$2Q_{\rm NLO}/(m_{\tilde{q}}+m_{\tilde{q}}) = [0.5;2.0]$. In this plot,
we set $m_{\tilde{q}}=m_{\tilde{g}}$.

\begin{figure}
 \begin{center}
  {\unitlength1cm
  \begin{picture}(16,14)
   \epsfig{file=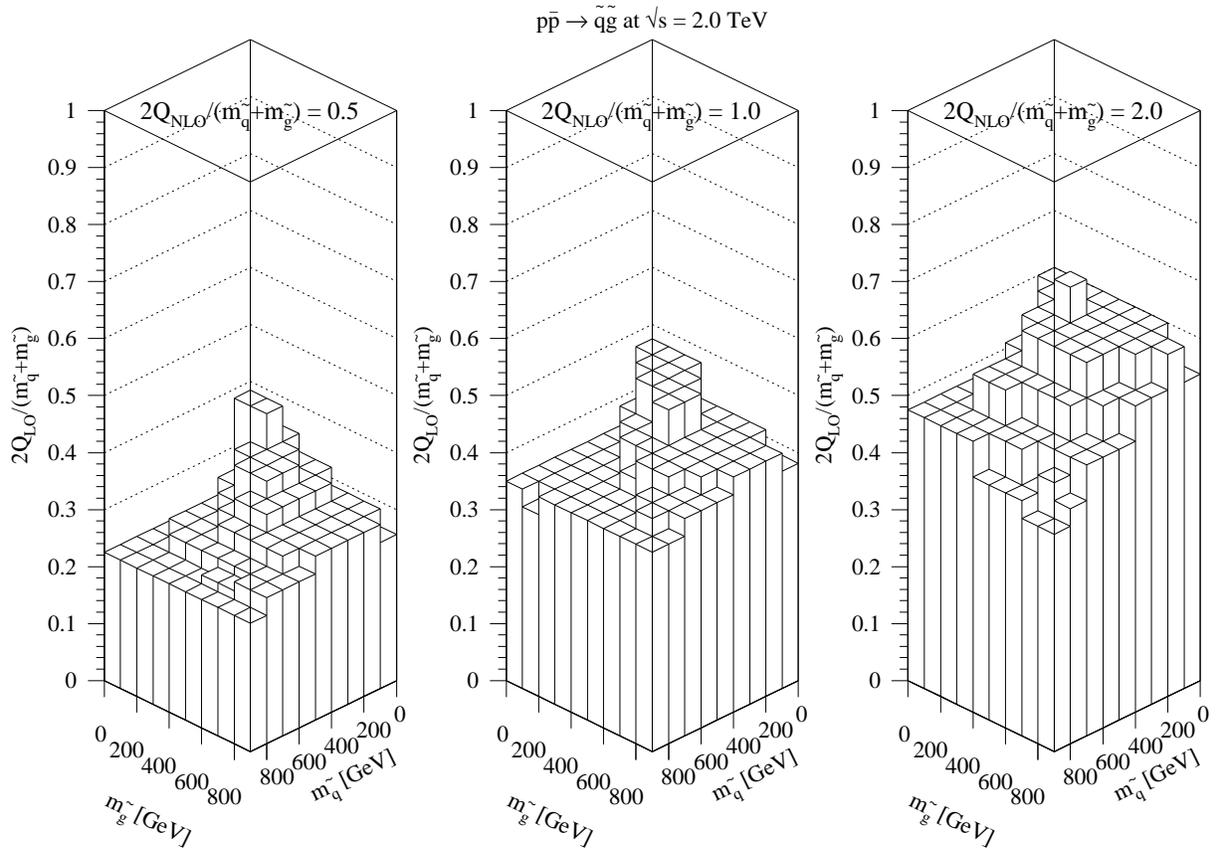,bbllx=535pt,bblly=90pt,bburx=85pt,bbury=715pt,%
           height=16cm,angle=270,clip=}
  \end{picture}}
 \end{center}
\caption{Optimal scale choices for associated production of a squark and 
gluino at
Run II of the Tevatron as a function of the squark mass $m_{\tilde{q}}$ and
the gluino mass $m_{\tilde{g}}$ for three different choices of
$2Q_{\rm NLO}/(m_{\tilde{q}}+m_{\tilde{g}}) = $
[0.5(left);1.0(middle);2.0(right)]. 
The dependence on $m_{\tilde{q}}$ and $m_{\tilde{g}}$
separately is much weaker than along the diagonal representing the average
squark and gluino mass $(m_{\tilde{q}}+m_{\tilde{g}})/2$ that 
sets the correct scale for the process.}
\label{fig6}
\end{figure}

The three-dimensional Fig.~\ref{fig6} shows the optimal scale choices for the 
same process as a function of both the squark mass $m_{\tilde{q}}$ and the 
gluino mass $m_{\tilde{g}}$. The dependence on $m_{\tilde{q}}$ and 
$m_{\tilde{g}}$ separately is much weaker than along the diagonal that 
represents the average squark and gluino mass 
$(m_{\tilde{q}}+m_{\tilde{g}})/2$.  This average would seem to be a 
preferable physical scale for the process instead of either of the two 
produced sparticle masses.

\begin{figure}
 \begin{center}
  {\unitlength1cm
  \begin{picture}(12,17)
   \epsfig{file=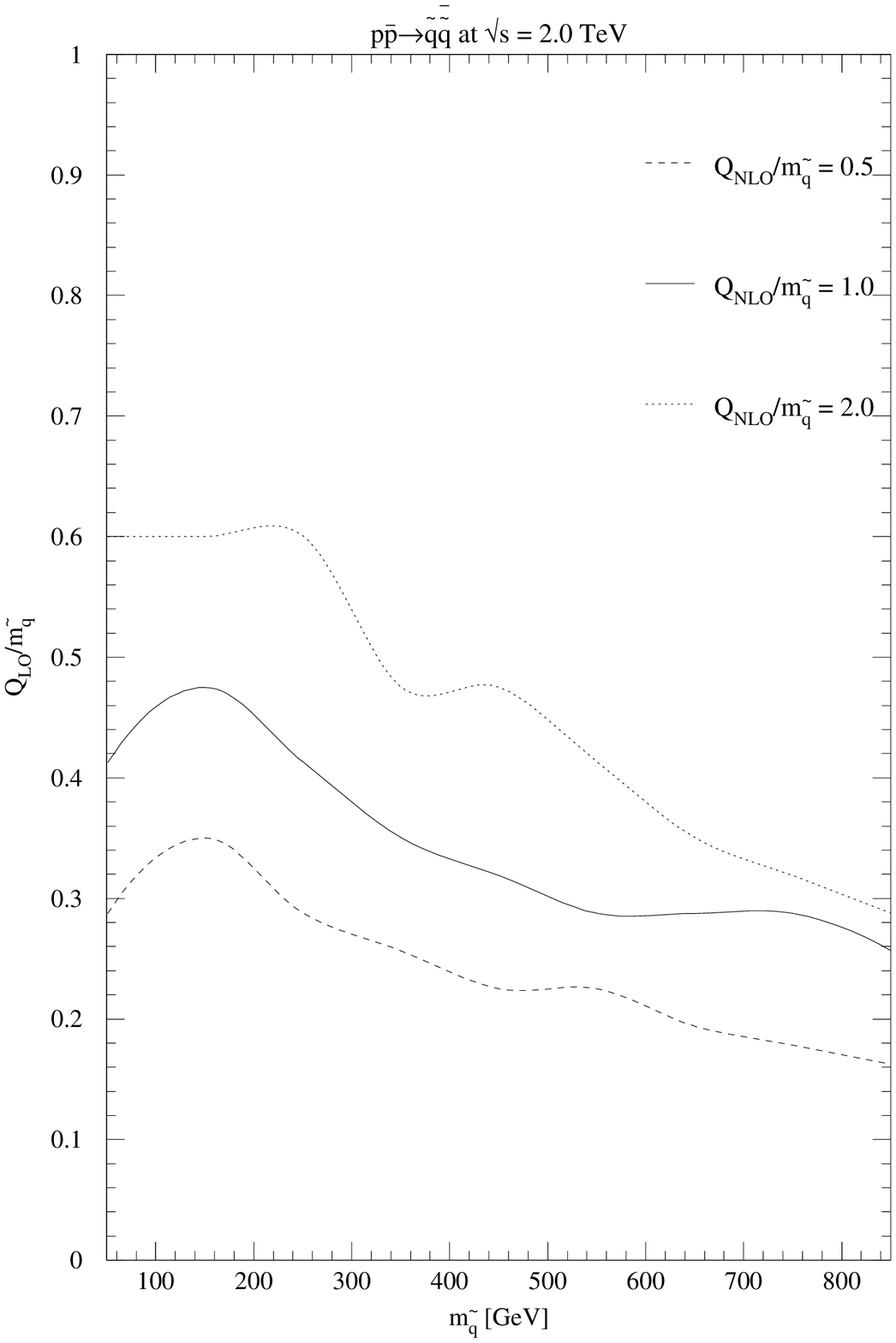,bbllx=60pt,bblly=95pt,bburx=485pt,bbury=730pt,%
           height=17cm,clip=}
  \end{picture}}
 \end{center}
\caption{Optimal scale choices for squark-antisquark production at Run II
of the Tevatron as a function of the squark mass $m_{\tilde{q}}$
with $m_{\tilde{g}}= 250$ GeV. For the
central (full) curve, the NLO scale has been chosen equal to the squark
mass. The dashed and dotted curves correspond to the remaining uncertainty at
NLO from a variation of $Q_{\rm NLO}/m_{\tilde{q}} = [0.5;2.0]$.}
\label{fig7}
\end{figure}

\begin{figure}
 \begin{center}
  {\unitlength1cm
  \begin{picture}(12,17)
   \epsfig{file=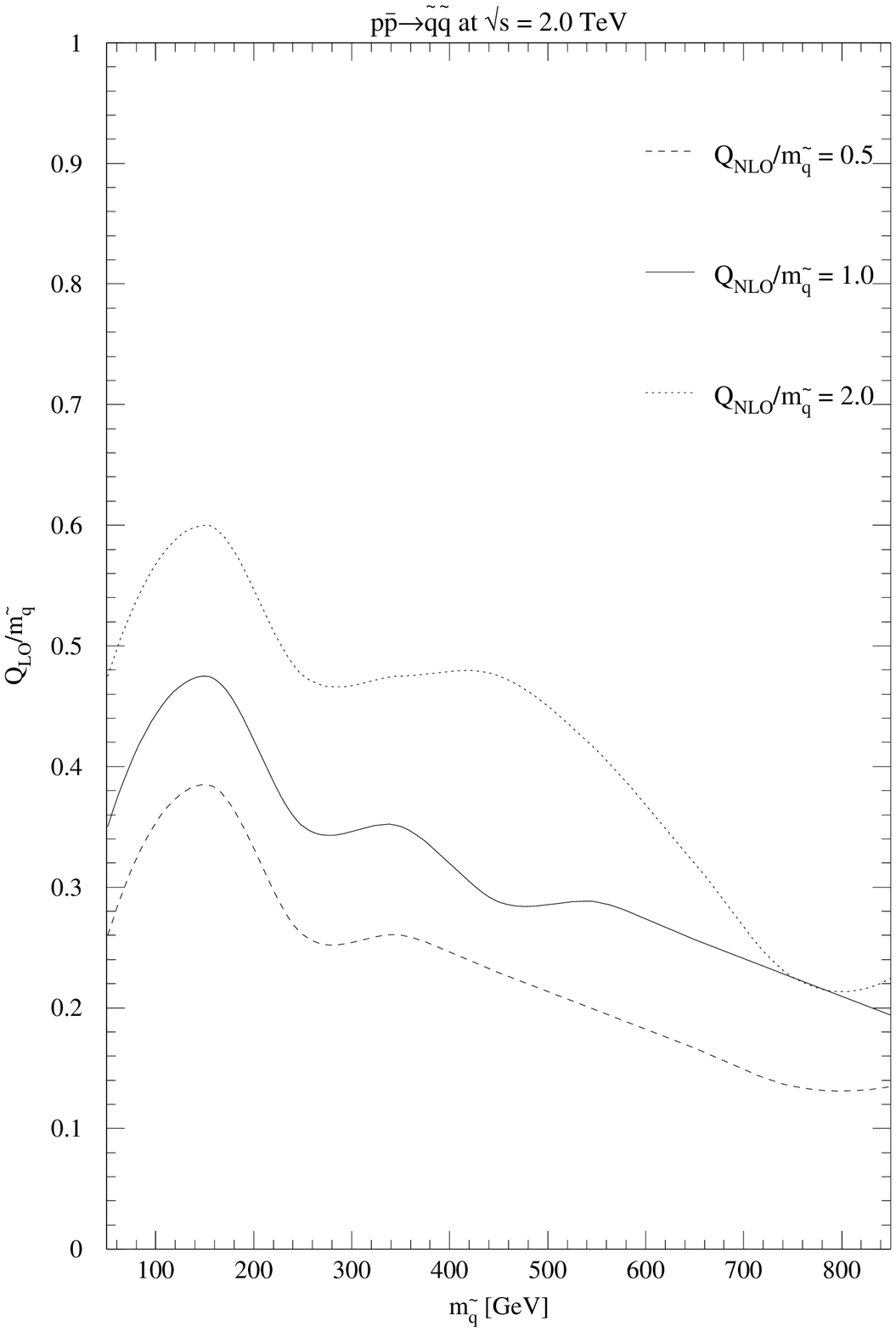,bbllx=60pt,bblly=95pt,bburx=485pt,bbury=730pt,%
           height=17cm,clip=}
  \end{picture}}
 \end{center}
\caption{Optimal scale choices for squark pair production at Run II
of the Tevatron as a function of the squark mass $m_{\tilde{q}}$
with $m_{\tilde{g}}= 250$ GeV. For the
central (full) curve, the NLO scale has been chosen equal to the squark
mass. The dashed and dotted curves correspond to the remaining uncertainty at
NLO from a variation of $Q_{\rm NLO}/m_{\tilde{q}} = [0.5;2.0]$.}
\label{fig8}
\end{figure}

\begin{figure}
 \begin{center}
  {\unitlength1cm
  \begin{picture}(12,17)
   \epsfig{file=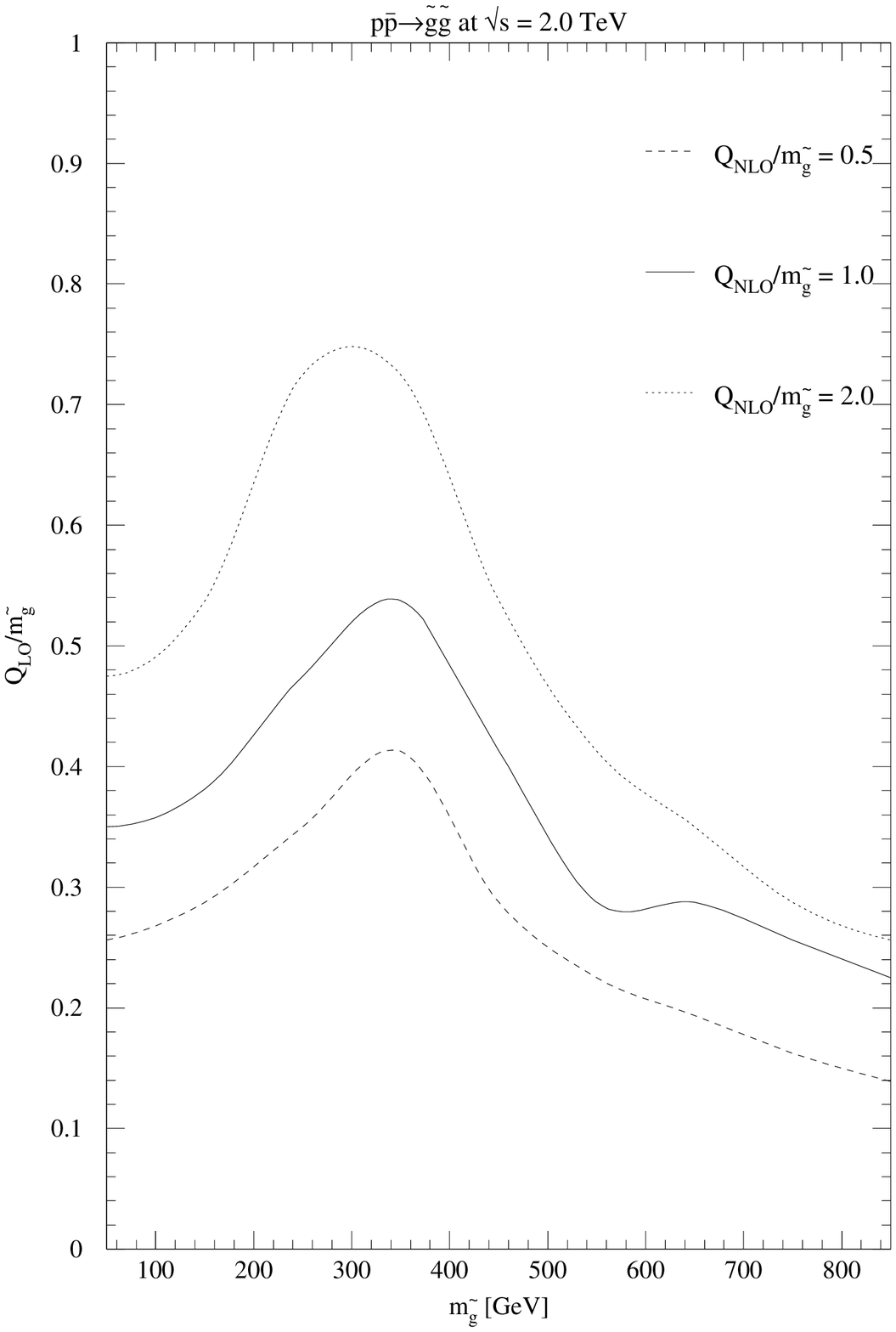,bbllx=60pt,bblly=95pt,bburx=485pt,bbury=730pt,%
           height=17cm,clip=}
  \end{picture}}
 \end{center}
\caption{Optimal scale choices for gluino pair production at Run II
of the Tevatron as a function of the gluino mass $m_{\tilde{g}}$
with the squark mass fixed at $m_{\tilde{q}}= 250$ GeV. For the
central (full) curve, the NLO scale has been chosen equal to the gluino
mass. The dashed and dotted curves correspond to the remaining uncertainty at
NLO from a variation of $Q_{\rm NLO}/m_{\tilde{g}} = [0.5;2.0]$. The
optimal scale shows a large sensitivity in the region where $m_{\tilde{g}}
= m_{\tilde{q}} = 250$~GeV.}
\label{fig9}
\end{figure}

\begin{figure}
 \begin{center}
  {\unitlength1cm
  \begin{picture}(12,17)
   \epsfig{file=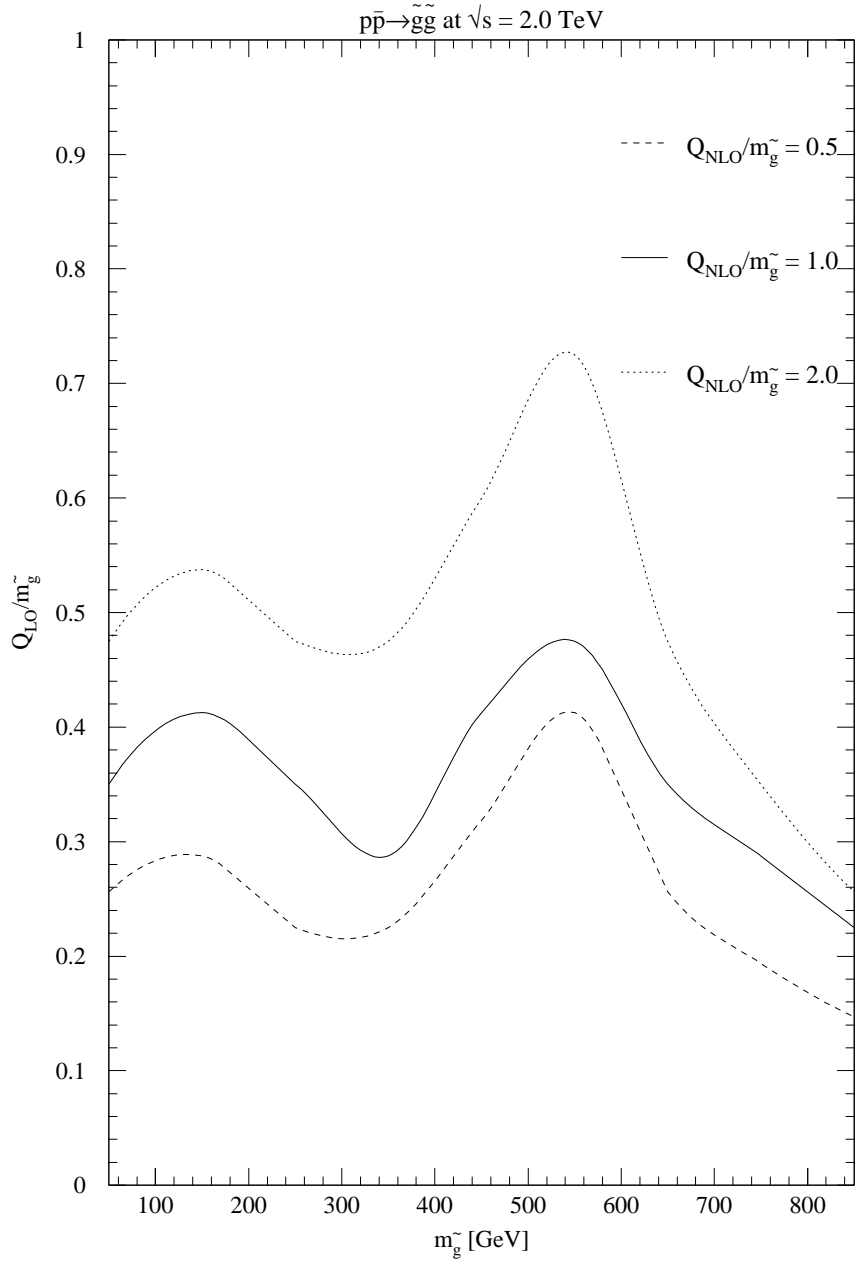,bbllx=60pt,bblly=95pt,bburx=485pt,bbury=730pt,%
           height=17cm,clip=}
  \end{picture}}
 \end{center}
\caption{Same as Fig.~\ref{fig9} for a squark mass of $m_{\tilde{q}} = 450$~GeV.}
%\ref{fig9} 
\label{fig10}
\end{figure}

\begin{figure}
 \begin{center}
  {\unitlength1cm
  \begin{picture}(16,14)
   \epsfig{file=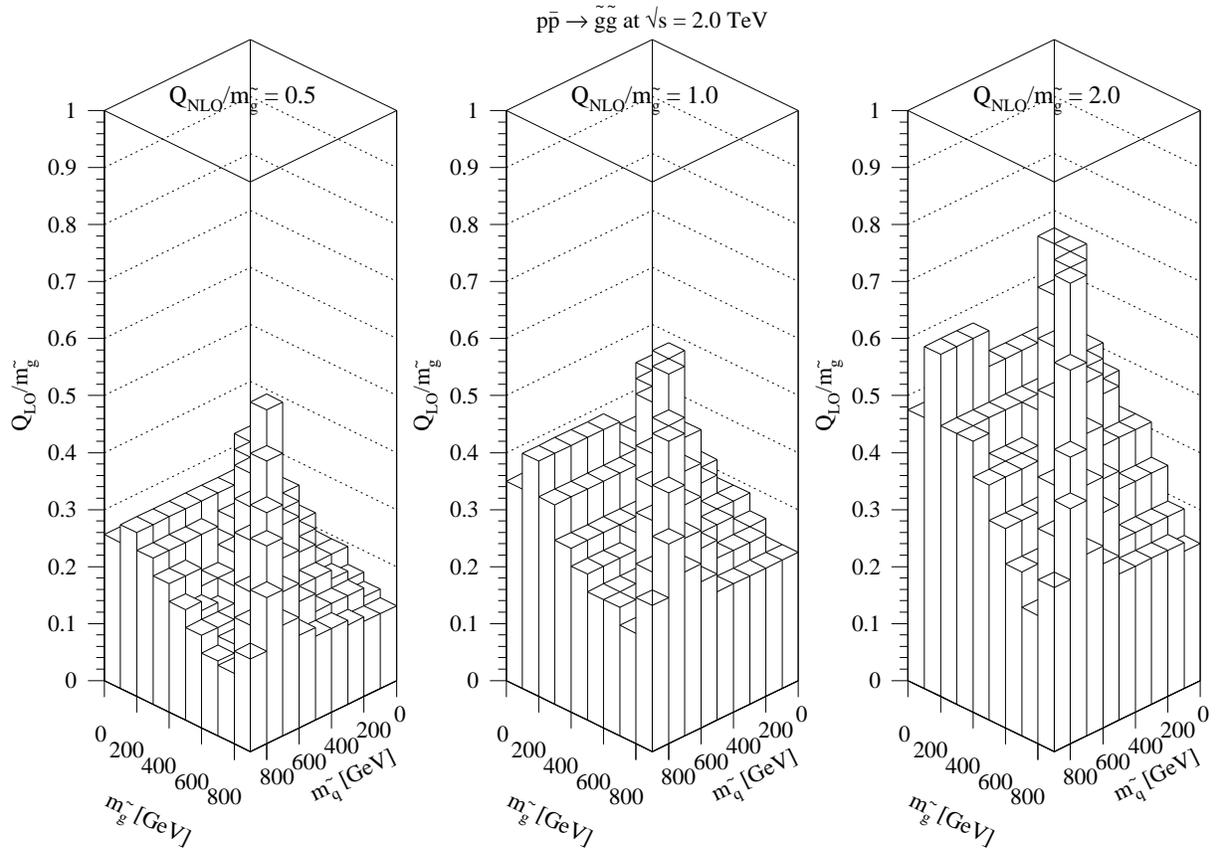,bbllx=535pt,bblly=90pt,bburx=85pt,bbury=715pt,%
           height=16cm,angle=270,clip=}
  \end{picture}}
 \end{center}
\caption{Optimal scale choices for gluino pair production at Run II
of the Tevatron as a function of the gluino mass $m_{\tilde{g}}$ and
the squark mass $m_{\tilde{q}}$ for three different choices of
$Q_{\rm NLO}/m_{\tilde{g}} = $ [0.5(left);1.0(middle);2.0(right)].
The optimal scale shows a large sensitivity along the diagonal where
$m_{\tilde{g}} = m_{\tilde{q}}$.}
\label{fig11}
\end{figure}

The squark mass dependence of the optimal scale choice for squark-antisquark
and squark pair production is plotted in Figs.~\ref{fig7} and \ref{fig8}.
For these processes, as well as for the associated production of squarks and
gluinos, the dependence on parameters other than the produced sparticle mass
is negligible.  On the other hand, in the case of gluino pair production 
the optimal scale shows a large sensitivity in the region where
$m_{\tilde{g}} = m_{\tilde{q}}$. The strong sensitivity is displayed in 
Figs.~\ref{fig9} and \ref{fig10}, where we set $m_{\tilde{q}} =$ 250 and 
450~GeV, respectively. The sensitivity arises from the fact that the LO 
Born cross section decreases with increasing squark mass in the region of
$m_{\tilde{g}} \sim m_{\tilde{q}}$, whereas the NLO virtual and soft
contributions increase.  A strong variation of the relative size of the two 
contributions results in that region.  Figure \ref{fig11} shows the
optimal scale choice for the same channel as a function of both the
gluino mass $m_{\tilde{g}}$ and the squark mass $m_{\tilde{q}}$.  The 
large sensitivity is apparent along the diagonal where 
$m_{\tilde{g}} = m_{\tilde{q}}$.  

\begin{figure}
 \begin{center}
  {\unitlength1cm
  \begin{picture}(12,17)
   \epsfig{file=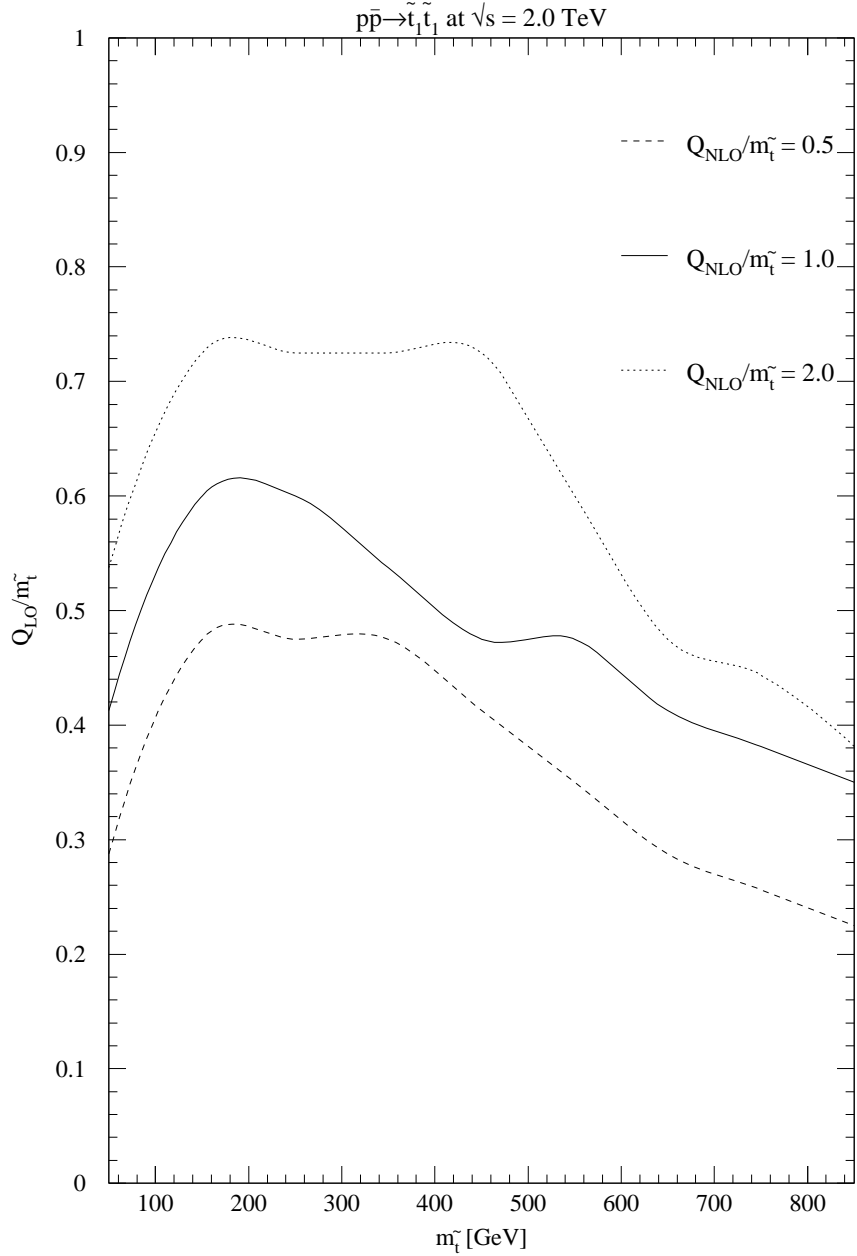,bbllx=60pt,bblly=95pt,bburx=485pt,bbury=730pt,%
           height=17cm,clip=}
  \end{picture}}
 \end{center}
\caption{Optimal scale choices for stop-antistop production at Run II
of the Tevatron as a function of the stop mass $m_{\tilde{t}_1}$. For the
central (full) curve, the NLO scale has been chosen equal to the stop
mass. The dashed and dotted curves correspond to the remaining uncertainty at
NLO from a variation of $Q_{\rm NLO}/m_{\tilde{t}_1} = [0.5;2.0]$.
The light squark mass, gluino mass, and mixing parameter
are the same as in Fig.~\ref{fig1e}.}
\label{fig12}
\end{figure}

The results for stop-antistop production are shown in Fig.~\ref{fig12}
as a function of the stop mass $m_{\tilde{t}_1}$. In this
case, dependence on the light squark and gluino masses is negligible
again.

For ease of implementation in Monte Carlo generators, we summarize our 
results in Table \ref{tab1}.   For the five different production channels 
we provide optimal values of the scale to be used in LO calculations in 
order to reproduce the values of the NLO cross sections obtained at 
three different values of the hard scale in the next-to-leading order 
calculations.  It is evident that one cannot adopt a single preferred LO scale 
for any one of the subprocess nor a single next-to-leading order scale, 
independent of sparticle mass,  unless one is prepared to bias the cross 
section.  The principal variation one sees is due to the dominant dependence 
of the cross sections on the produced sparticle mass, and the optimal 
scale generally decreases with increasing sparticle mass.

%%%%%%%%%%%%%% End of Section IV %%%%%%%%%%%%%%%%%%%%%%%%%%%%%%%%%%%%%%%%%

%%%%%%%%%%%%%% Begin Section V %%%%%%%%%%%%%%%%%%%%%%%%%%%%%%%%%%%%%%%%
\section{Summary}
\label{sec:5}

In this paper, we present next-to-leading order
cross sections for squark and gluino production at a center-of-mass energy 
of $\sqrt{s_H} = 2$ TeV.  The hard scale dependence of the cross section at 
leading order in perturbative QCD is reduced at NLO but not entirely absent.   
For three choices of the scale used in the next-to-leading order calculations, 
we define optimal scales that can be used in leading order Monte Carlo 
simulations.  If these optimal 
scales are used, the size of the next-to-leading order cross section is 
reproduced.  The values of the optimal scales depend mainly on the produced 
sparticle mass, with the exception of the case of gluino pair production 
where sensitivity in the region near $m_{\tilde{g}}=m_{\tilde{q}}$ requires 
special attention.  One may be tempted to use a single scale in all processes 
and for all masses, but the price is to bias the cross section determination. 
Our results allow one to select hard scales in LO calculations so that the 
resulting cross sections will reproduce 
a range of next-to-leading order cross sections, from optimistic to pessimistic 
values.

In this paper, cross sections are for inclusive yields, integrated over 
all transverse momenta and rapidities.  In the search for supersymmetric 
states, a selection on transverse momentum will normally be applied in 
order to improve the signal to background conditions.  Our analysis can also 
be carried out with similar selections.  
%Those results will be reported elsewhere.  
A tabulation of cross sections for various squark and gluino masses is available
upon request.

There is little doubt that the reliability of Monte Carlo codes is an 
important issue.  All current exclusion limits on masses of sparticles are 
based on Monte Carlo simulations.  However, not enough is known about the 
systematic uncertainties associated with the use of Monte Carlo simulations, 
especially in view of the fact that signals for sparticles are sought in 
relatively small corners of phase space.  A change in the hard scale from 
the oft assumed value of $Q/{\tilde{m}}$ approximately 1.0 to $Q/{\tilde{m}}$
in the interval 0.2 to 0.6, as we determine, will change the transverse 
momentum distributions of sparticles in the final state, making their 
detection either more or less probable.  To find out how much change results, 
one must compare results obtained from runs of the Monte Carlo codes with the 
different values of the scale.  The change will provide a new measure of the 
theoretical systematic uncertainty.  

%%%%%%%%%%%%%% Begin Acknowledgment %%%%%%%%%%%%%%%%%%%%%%%%%%%%%%%%%%%%
\section*{Acknowledgments}

Work in the High Energy Physics Division at Argonne National Laboratory
is supported by the U.S. Department of Energy, Division of High Energy
Physics, Contract W-31-109-ENG-38.  We have benefitted from communications with 
H. Baer, M. Spira, and C. Wagner, and are grateful to M. Spira for providing a copy
of his numerical code. 
%%%%%%%%%%%%%% End of Acknowledgment %%%%%%%%%%%%%%%%%%%%%%%%%%%%%%%%%%%

%%%%%%%%%%%%%% Begin References %%%%%%%%%%%%%%%%%%%%%%%%%%%%%%%%%%%%%%%%

%%%%%%%%%%%%%% End of References %%%%%%%%%%%%%%%%%%%%%%%%%%%%%%%%%%%%%%%

%%%%%%%%%%%%%% Begin Tables %%%%%%%%%%%%%%%%%%%%%%%%%%%%%%%%%%%%%%%%%%%%

 \begin{table}[h]
 \begin{tabular}{llll}
 \hline
  Channel & $Q_{\rm NLO}/\tilde{m} = 0.5$ & $Q_{\rm NLO}/\tilde{m} = 1.0$ &
  $Q_{\rm NLO}/\tilde{m} = 2.0$ \\
 \hline
  $\tilde{q}\bar{\tilde{q}}    $ & 0.35 - 0.15 & 0.45 - 0.25 & 0.60 - 0.30 \\
  $\tilde{q}     \tilde{q}     $ & 0.40 - 0.15 & 0.45 - 0.20 & 0.60 - 0.20 \\
  $\tilde{g}     \tilde{g}     $ & 0.45 - 0.15 & 0.50 - 0.20 & 0.75 - 0.25 \\
  $\tilde{q}     \tilde{g}     $ & 0.40 - 0.20 & 0.50 - 0.35 & 0.60 - 0.40 \\
  $\tilde{t}_1\bar{\tilde{t}}_1$ & 0.50 - 0.20 & 0.60 - 0.35 & 0.70 - 0.40 \\
 \hline
 \end{tabular}
 \caption{Optimal scale choices for the production of squarks and gluinos
 at the Tevatron Run II for five different production channels and three
 different choices of the next-to-leading order scale. The variation is
 due to the dominant dependence of the optimal scale on the produced sparticle 
 mass over the range $\tilde{m}$ from 50 to 850~GeV.  The optimal scale  
 generally decreases with increasing sparticle mass.}
 \label{tab1}
 \end{table}

%%%%%%%%%%%%%% End of Tables %%%%%%%%%%%%%%%%%%%%%%%%%%%%%%%%%%%%%%%%%%%

%%%%%%%%%%%%%% Begin Figure Captions %%%%%%%%%%%%%%%%%%%%%%%%%%%%%%%%%%%

%%%%%%%%%%%%%% End of Figure Captions %%%%%%%%%%%%%%%%%%%%%%%%%%%%%%%%%%

\end{document}